# The Influence of Interchain Coupling on Intramolecular Oscillation Mobility in Coupled Macromolecular Chains: The case of Coplanar Parallel Chains

D. Čevizović[1,b)], S. Petković[1,c)], S. Galović[1,d)], A. Chizhov[2,e)] and A. Reshetnyak[3,a)]

[1]*University of Belgrade, "Vinča" Institute of Nuclear sciences, Laboratory for Theoretical and Condensed Matter Physics, P.O. BOX 522, 11001, Belgrade, Serbia .*
[2]*Joint Institute for Nuclear Research, Bogoliubov Laboratory of Theoretical Physics, Joliot-Curie 6, 141980 Dubna, Moscow region, Russia.*
[3]*Institute of Strength Physics and Materials Science SB RAS, 2/4, pr. Akademicheskii, Tomsk, 634021, Russia*

[a)]Corresponding author: [)]reshet@ispms.tsc.ru
[b)]cevizd@vinca.rs
[c)]stanmark@gmail.com
[d)]bobagal@vinca.rs
[e)]chizhov@theor.jinr.ru

**Abstract.** We enlarge our results from the study of the hopping mechanism of the oscillation excitation transport in 1D model of one biologica-likel macromolecular chain [6], [7] to the case of a system composed from two 1D parallel macromolecular chains with consideration of the properties of intramolecular oscillation excitations. We suppose, that due to the exciton interaction with thermal oscillation (generated by mechanical phonon subsystem) of structural elements (consisting of the peptide group) of the chains, the exciton becomes by self trapped and forms the polaron state. We suggest a model which generalizes the modified Holstein polaron model [12] to the case of two macromolecular chains and find that because of the interchain coupling, the exciton energy band is splitted into two subbands. The hopping process of exciton migration along the macromolecular chains is studied in dependence of system parameters and temperature. We pay an special attention to the temperature range (near T=300 K) in which living cells operate. It is found that for the certain values of the system parameters there exists the abrupt change of the exciton migration nature from practically free (light) exciton motion to an immobile (heavy, dressed by phonon cloud) quasiparticle We discuss an application of the obtained results to the exciton transport both within deoxyribonucleic acid molecule and in the 2D polymer films organized from such macromolecular chains.

## INTRODUCTION

The DNA macromolecule, polysaccharides and some other biological structures being consisting of several parallel macromolecular chains (MCs) have an important role in the process of charge,energy or bio-information transferring in the living cells. In turn, because of theirs special peculiriaties, like enough large length near 10-$10^2$mkm, stability to the temparature changing near 100K-400K and superelasticity, they are able to realize the problem of miniaturizing of microelectronic and optoelectronic devices that is providing a renewed interst for an application of such structures in construction of nanocrystals, nanowires and molecular circuits [1,2,3]. An efficient application of these materials must have a completely strong understanding of the process of charge and energy transport along macromolecule starting from the quantum level. The last fact esspecially based on the correct theoretic models about the energy transport at the distances which are comparable to the length of a macromolecule (long distance transport, about 0,01m). Note, there exists the strong arguments that the process of hydrolise of adenosine triphosphate (ATP) may be considered as the main source of bioenergy being necessary for functioning of the cell metabolism. In this process. one arises the energy quanta being quite enough to excite the intramolecular

(IMO) excitation (vibron) on the peptide bond (Amide I or CO quanta of stretching). However, in spite of many efforts the mechanism which provides the migration of such IMO quanta to long distance still remains not clear.

The main aim of the paper is to study the process of IMO excitation migration in the system which consists of two coplanar and parallel macromolecular chains. Among the typical examples of such structure we can specify the ACN macromolecule in which the two polypeptide chains are embedded and form two chain structure (FIGURE.1).

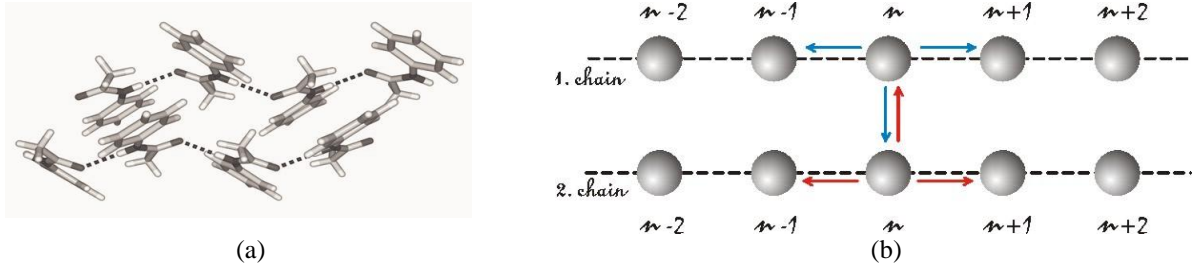

(a)  (b)

**FIGURE 1.** The real structure of the part of crystaline acetanilide macromolecular chainis is shown on the left (a); whereas on the right (b) it is demonstrated its schematic structure. The possible vibron hopping transitions are shown by the arrows

In order to investigate the IMO migration along the macromolecular spine, we suppose that IMO excitation can be localized into self-trapped (ST) state, due to the interaction with thermal oscillation of the macromolecular chains. In this case IMO excitation should form the (partially) dressed (by phonons) quasiparticle. One should be noted, the results of the incoherent neutron scattering expreriments indicates that acoustic phonon modes are not involved in the process of vibron ST organization in the MCs [4]. Having this fact in mind, we confine our research to the case of vibron interaction with non-dispersive optic phonon modes, only. As the theoretical basis of our study, we will use improved Holstein molecular crystal model [5,6].

The paper is organized as follows. In the second Section we introduce the theoretical model being based on the new form of the quantum Hamiltonian. We analyse there the influence of the system temperature and basic system parameters to the process of quasiparticle dressing and explicitly calculate the probability of the quasiparticle migration from one to its neighbor macromolecular site, in the dependence of the system parameters and temperature. In the final Section we determine, as it was done for the case of single MCs [6,7], a region in the space of the system parameter where IMO dressing, as well as its mobility undergo the drastic changes.

## THE THEORETICAL MODEL

For the purpose of study the properties of the single vibron excitation in the system consisting of two parallel macromolecular chains, we suppose that because of the vibron–phonon interaction, the vibron excitation becomes self-trapped and in fact forms a dressed quasiparticle. As a theoretical framework, we modify the Holstein molecular crystal model [5,6] and adopt several natural assumptions:

1. We consider a single vibron (not double) excitation, excited on the $n$-th structural element of a $j$-th chain for $j=1,2$ and suppose that the vibron properties are determined by the thermal oscillations of the MCs
2. The thermal oscillations of any single MC from double chain do not influence on a thermal oscillation of the another one (i.e. we neglect by the interaction between phonon subsystems that belongs to the different chains). As the consequence, the single chaines in the double MC looks as independent and therefore may be considered as identical ones leading to equality of, the phonon spectra that belong to different chains
3. The vibron excited on the particular MC must interact with only phonons that belong to the same MC.
4. We suggest because of the energy of the dipole-dipole interaction between the first-neighboring peptide groups is larger than the energy of the dipole-dipole interaction between second-neighboring groupes [8], that a vibron can only migrate along the macromolecule from $n$-th to ($n\pm1$)-th structural element.

The Hamiltonian of the system of 2 parallel and unstaggered from each other MCs (FIGURE.1b) is defined as:

$$H = E_0 \sum_{n,j=1,2} A^+_{j,n} A_{j,n} - J \sum_{n,j=1,2} A^+_{j,n}\left(A_{j,n-1} + A_{j,n+1}\right) + L \sum_{n,j=1,2}\left(A^+_{1,n} A_{2,n} + A^+_{2,n} A_{1,n}\right) + \sum_{j,q} \hbar\omega_q B^+_{j,q} B_{j,q}$$
$$+ \frac{1}{\sqrt{N}} \sum_{n,q,j=1,2} F_q e^{iqnR_0} A^+_{j,n} A_{j,n}(B_{j,q} + B^+_{j,-q}) , \quad (1)$$

where $E_0$ appears by the vibron excitation energy on an particular MC site, $A^+_{j,n}$ ($A_{j,n}$) are vibron quasiparticle creation (annihilation) operators on $n$-th site of $j$-th MC, $J$ is energy of dipole-dipole interaction of neighboring structure elements that belongs to the same chain, while $L$ corresponds to the energy of dipole-dipole interaction between structure elements that belong to different chains. The operators $B^+_{j,q}$ ($B_{j,q}$) in (1) are the creation (annihilation) operators of phonon quanta with frequency $\omega_q$, wavenumber $q$ which belong to $j$-th chain. In the case of the exciton interaction with the optical phonon modes, the exciton-phonon coupling parameter is: $F_q = \chi_{hb}\sqrt{\frac{\hbar}{2M\omega_0}}$, (with $\chi_{hb}$ being by the coupling constant, $M$ is a mass of molecular group and $\omega_q = \omega_0$). $R_0$ is distance between two neighboring structure elements, placed on the same MC. In the standard way the transition to the polaron picture is realized by use of the modified Lang-Firsov unitary transformation [5,6,9,10] with operator: $U = U_1 \cdot U_2$, where $U_j = e^{-\sum_n \sigma_{j,n} A^+_{j,n} A_{j,n}}$, and $\sigma_{j,n} = \frac{1}{\sqrt{N}}\sum_q f_{j,q} e^{-iqnR_0}(B_{j,q} - B^+_{j,-q})$ but now for double MC. Then, we use the variational approach with only one single variational parameter $\delta \in [0,1]$ [9] introduced as follows $f_q = \delta \frac{F^*_q}{\hbar\omega_0}$. The influence of the thermal fluctuations on the vibron properties are accounted by the averaging of the transformed Hamiltonian over the phonon subsystem [10,11]. Doing so, we find the form

$$H_{vib} = \sum_{k,j=1,2} E(k) a^+_{j,k} a_{j,k} + \lambda_0 \sum_k \left(a^+_{1,k} a_{2,k} + a^+_{2,k} a_{1,k}\right), \qquad (2)$$

of the dressed vibron Hamiltonian $H_{vib}$, where $a^+_{j,k}$ ($a_{j,k}$) denotes the creation (annihilation) operator of the dressed vibrons, $\tau = k_B T/\hbar\omega_0$ is normalized temperature, $\lambda_0 = L e^{-\delta^2 S \operatorname{cth}(1/2\tau)}$, and

$$E(k) = E_0 + \delta(\delta - 2)E_B - 2Je^{-\delta^2 S \operatorname{cth}(1/2\tau)}\cos(kR_0), \quad E_B = F^2/\hbar\omega_0, \quad S = E_B/\hbar\omega_0 \qquad (3)$$

appear respectively by the polaron band energy, small polaron binding energy, and coupling constant. The diagonalization of the hamiltonian (2) can be done by means of the unitary transformation $\alpha_{\pm,k} = \frac{1}{\sqrt{2}}(a_{1,k} \pm a_{2,k})$. As the result the vibron hamiltonian becomes by diagonalized operator

$$H_{vib} = \sum_k (E_-(k)\alpha^+_{-,k}\alpha_{-,k} + E_+(k)\alpha^+_{+,k}\alpha_{+,k}) \qquad (4)$$

while polaron band becomes splitted into two subbands: $E_\pm(k) = E_0 + \delta(\delta-2)E_B - 2Je^{-\delta^2 S \operatorname{cth}(1/2\tau)}\cos(kR_0) \pm \lambda_0$. The optimal state of the system is determined by the minimum of the free energy of whole system. Therefore, the value of the variational parameter should follow from the conditions $dF_B/d\delta = 0$ and $d^2F_B/d\delta^2 > 0$, where $F_B$ is the upper bound of the system free energy, determined by the Bogoliubov theorem [6,10]. The probability of the vibron transition from one to another nearest neighbour structural elemnet, along the macromolecule spine is determined according to [12] by the relation:

$$w(T) = \frac{\omega_0 B^2}{4\delta}\sqrt{\pi \sinh(1/2\tau)}\, e^{-\delta^2 S \tanh(1/4\tau)} \qquad (5)$$

with $B = 2J/k_B T$ being by a so called *adiabatic parameter*.

# OBTAINED RESULTS AND DISCUSSION

To study the influence of the interchain coupling to the vibron hopping process in the double MC composed from two unstaggered single MCs, we have firstly considered the model described by new Hamiltonian (1) and calculated the probability of the vibron transition along the macromolecule spine for several values of interchain coupling parameters $g = J/L$ which is depending on the coupling constant $S$ and adiabatic parameter $B$ in (5) and evaluated it at the room temperature 300 K. The results of graphical treatment are presented on the FIGURE 2.

From the presented results it follows that the space of the parameters for the system (determined by $S$ and $B$) is divided on two subspaces: one where the probability of the vibron transfer seems by relatively high and when vibron moves as a practically free quasiparticle (in this subspace $B$ is high while $S$ is small), and subspace where the vibron has small mobility (i.e. where vibron becomes immobile and remains localized at a particular site of MC).. The latter case is characterized by the small values of $B$. These regions are divided by the subset of critical points for which we can undergo a drastic change of the probability of the vibron transfer. It is obvious, the region of the critical subset strongly depends on the interchain coupling, and for sufficiently large interchain coupling it can enter into region of system parameters that correspond to real protein molecules (as it follows from the FIGURE 2.(c)) with the first point where probability of the vibron transfer gets in the zone of small values of $B$ and $S$). We intend to develop the models to consider the exciton transport in DNA molecule and in 2D polymer films from such MCs.

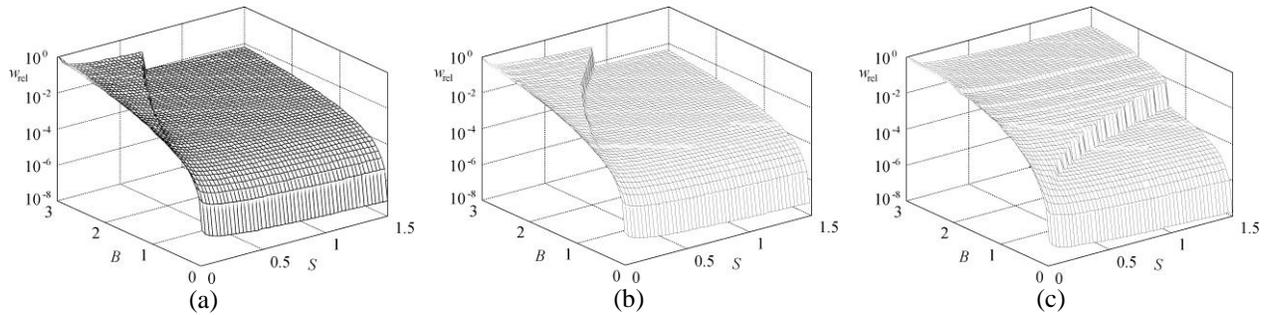

**FIGURE 2**. The dependence of the migration probability on the parameters $S$, $B$ for 3 different values of the interchain couplings: $g$=0.5 (a); $g$=1.0 (b) and $g$=5.0 (c) for $T$=300K. The literature values [9] of the basic physical parameters of ACN suggest that the values of $S$ and $B$ that corresponds to ST state of the Amide-I quanta in ACN are $S \sim 0.25$ and $B \sim 0.16$.

# ACKNOWLEDGMENTS

The work was supported by the Serbian Ministry of Education and Science, under Contract Nos. III-45010, III-45005, and by the Project within the Cooperation Agreement between the JINR, Dubna, Russian Federation and Ministry of Education and Science of Republic of Serbia. The work of R.A.A.was partially supported by the grant of Leading Scientific Schools of the Russian Federation under Project No. 88.2014.2.